# A Bibliometric Analysis of Highly Cited Artificial Intelligence Publications in Science Citation Index Expanded

Yuh-Shan Ho[*] ; Juan-Jose Prieto-Gutierrez[**]

[*] CT HO Trend
3F.-7, No. 1, Fuxing N. Rd., Songshan Dist., Taipei City 105611, Taiwan
Orcid: 0000-0002-2557-8736
E-mail: dr.yuhshanho@gmail.com

[**] Library and Information Department, Complutense University of Madrid, Spain
Orcid: 0000-0002-1730-8621
E-mail: jjpg@ucm.es

## Abstract

This study aimed to identify and analyze the characteristics of highly cited publications in the field of artificial intelligence within the Science Citation Index Expanded (SCI-EXPANDED) from 1991 to 2022. The assessment focused on documents that garnered 100 citations or more from the Web of Science Core Collection, spanning from their publication date to the end of 2022. Various aspects of these documents were analyzed, encompassing document types, the distribution of annual production, the average number of citations per publication, Web of Science categories, and journals. Moreover, the publication performance of countries, institutions, and authors underwent evaluation through six publication indicators and associated citation metrics. To facilitate a comprehensive comparison of the authors' research performance, the $Y$-index was employed. The outcomes of the analysis revealed that a majority of the highly cited articles were published within the Web of Science categories of 'artificial intelligence computer science' and 'electrical and electronic engineering'. Notably, the United States exhibited dominance across all six publication indicators. Within the realm of average citations per publication, the United Kingdom emerged as a leader for independent articles, first-author articles, and corresponding-author articles. Exceptionally, the Chinese Academy of Sciences in China and the Massachusetts Institute of Technology (MIT) in the USA, contributed significantly. The significant impact of highly cited articles extended to the output of Stanford University in the USA. B.L. Bassler published the most highly cited articles. Upon employing the $Y$-index analysis, J.E.P. Santos was identified as having the highest potential for publication. In addition to the primary analysis, this study also presented nine classic articles that have left an indelible mark on artificial intelligence research.

# INTRODUCTION AND LITERATURE REVIEW

There is no doubt that society is increasingly moving towards widespread technification. The trend is inevitable and advances in technology have been constant over the last few years (Deguchi et al., 2020). Those with widespread application of information and communication technologies, such as artificial intelligence (AI), have been among the latest to become popular on a large scale, and it is even argued that the impacts of AI will be greater than those caused by the industrial revolutions of the 20th century (Miailhe and Lannquist, 2018). Although uncertainty is wide, economic theory warns that technological progress is likely to create both winners and losers (Korinek and Stiglitz, 2019).

Going back to the beginning of technological progress in AI, the origins of this progress officially date back to the mid-19th century. Specifically, to the year 1843 when British mathematician Ada Lovelace published the first computer algorithm in history aimed at a machine (Aiello, 2016) and, years later, Charles Babbage designed his "Analytical Engine or Engine", thanks to which she can be considered the first programmer (Babbage, 1973). But it was in 1956 that the term "artificial intelligence" was proposed at a conference at Dartmouth University, where a new topic of research was initiated (Hamet and Tremblay, 2017).

This, sometimes called machine intelligence or designed intelligence (Heift, 1998) is difficult to define, although over the years it can be defined as the intelligence demonstrated by machines in relation to the natural intelligence of humans and animals. Thus, it is the action where computers simulate the intelligent activities and behaviours of humans, and, as proposed in 1956, AI can produce intelligent machines that can reason, learn, communicate, plan, move, operate objects and solve problems (McCarthy, 1956).

IA has emerged as a possible solution to multiple obstacles and problems in all sectors such as pharmacology (Korteling et al., 2021), healthcare (Malik et al., 2019), science writing (Salvagno et al., 2023), education (Chen, et al, 2020), architecture and construction (Debrah et al., 2022), and so on. Artificial intelligence shows a growing and extensive presence in academic discourse (Riahi et al., 2021), being exploited by bibliometric analysis techniques to identify patterns of behaviour and trends of the emerging topic (AI) in a given field, such as those dedicated to AI publications related to healthcare (Guo et al., 2020), to observe the progress of AI in the field of tourism (Kirtil and Aşkun, 2021), to better understand the evolution of AI research in renewable energy (Zhang et al., 2022), behaviour in e-commerce (Bawack et al., 2022), application in the maritime industry sector (Munim et al., 2020), etc. Therefore, a very polarized perspective on a given area or category is available.

Of the few existing global researches, there is one (Niu et al., 2016), which is not current, where through database of the Science Citation Index Expanded (SCI-EXPANDED) and Conference Proceedings Citation Index-Science (CPCI-S) it is concluded that scientific production increases widely from the 1990s onwards, that computer science and engineering were the most active subjects in AI and the United States was the country that published the most on the subject.

Another more recent study (Ho and Wang, 2020) indicates that articles related to artificial intelligence were published in a wide range of Web of Science journals and categories. The United States is the country that publishes the most (20%) followed by China (13%) (Prieto-Gutierrez et al., 2023). The results of the word cluster analysis showed that the most popular topics were models, neural networks, learning and prediction. In this context, this study aims to identify, analyse, and characterized the impact of highly cited publications on artificial intelligence in the Science Citation Index Expanded.

This is the first kind of study, which would provide insight into characteristics of the highly cited articles.

## METHODOLOGY

The data reported in this study were retrieved from the online version of the Science Citation Index EXAPNDED (SCI-EXPANDED), the Clarivate Analytics Web of Science Core Collection database (data updated on 28 June 2023). The 2022 journal's impact factor ($IF_{2022}$) was reported in the Journal Citation Reports (JCR) on 28 June 2023. According to the definition of the journal's impact factor, Chiu and Ho (2021) recommended to search documents published in 2022 from SCI-EXPANDED after $IF_{2022}$ was presented (Chiu and Ho, 2021). Quotation marks (" ") and Boolean operator "or" were used which ensured the appearance of at least one search keyword in the terms of TOPIC (title, abstract, author keywords, and *Keywords Plus*) from 1991 to 2022. The search keywords used were: "artificial intelligence", "artificial intelligent", and "artificially intelligent". To have more accurate analysis results, uncommon terms: "artificial intelligences", "artificially-intelligence", "artificial intelligenced", "artificial intelligencer"; misspelling terms: "artificia intelligence", "artificial intelligencex", "artificial intelligenge", "artificial intellingence", "artificially intelligen", and "artificical intelligence"; and some terms missed spaces in the database: "artificial intelligencebased", "artificial intelligenceassisted", "artificial intelligencefind", and "artificial intelligencefor" in SCI-EXPANDED were also considered. Furthermore, abbreviation "AI" of "artificial intelligence" was also considered. However, publications contain "AI" which are irrelevant to "artificial intelligence", for example, "% ai", "abundance index (AI", "adequacy index (AI)", "AI alloy", "AI(X)", "ai/ha", "AI-AII", "AI-BSA", "Ai-Dere", "AI-Li", "aluminum (AI)", "amnioinfusion (AI)", "angiotensin I (AI)", "aortic insufficiency (AI)", "apnea index (AI)", "apnoea index (AI)", "apo-AI", "apoproteins (apo) AI", "artificial insemination (AI)", "associative ionization (AI)", "auditory cortex (AI)", "auditory field (AI)", "autoimmune (AI)", "avian influenza (AI)", "element-of Ai", "g ai", "kg ai", and "Kozlowski ct ai" were discard.

The total number of citations from Web of Science Core Collection received since publication year till the end of the most recent year of 2022 ($TC_{2022}$) (Wang et al., 2011) was used. Using $TC_{2022}$ is advantageous owing to its invariability and ensured repeatability as compared to the number of citations from the Web of Science Core Collection directly (Ho and Hartley, 2016a). Publications with $TC_{2022}$ of 100 or more were selected as highly cited publications (Ho, 2014b). A total of 3,182 highly cited artificial intelligence documents were found in SCI-EXPANDED from 1991 to 2022. *Keywords Plus* supplies additional search terms extracted from the titles of articles cited by authors in their bibliographies and footnotes in the Institute of Science Information (ISI) (now Clarivate Analytics) database, and substantially augments title-word and author-keyword indexing (Garfield, 1990). It was pointed out that documents only searched out by *Keywords Plus* are irrelevant to the search topic (Fu and Ho, 2015). Ho's group firstly proposed the 'front page' as a filter including the article title, abstract, and author keywords (Wang and Ho, 2011). Finally, 3,044 documents (96% of 3,182 documents) including search keywords in their 'front page' were defined as highly cited artificial intelligence research publications.

The full record in SCI-EXPANDED and the number of citations in each year for each document were checked and downloaded into Excel Microsoft 365, and additional coding was manually performed (Li and Ho, 2008; Al-Moraissi et al., 2023). The functions in the Excel Microsoft 365, for example, Counta, Concatenate, Filter, Match, Vlookup, Proper, Rank, Replace, Freeze Panes, Sort, Sum, and Len were applied (Al-Moraissi et al., 2023). The journal impact factors ($IF_{2022}$) were taken from the Journal Citation Reports (JCR) published in 2022.

In the SCI-EXPANDED database, the corresponding author is labelled as reprint author, but in this study, we used the term corresponding author (Chiu and Ho, 2007). Single authors in articles with unspecified authorship were both the first as well as corresponding authors (Ho, 2014a). The single institution in articles with unspecified corresponding institutions was both the first as well as corresponding-author institutions (Ho, 2014a). Similarly, in a single-country article, the country is classified as the first as well as the corresponding-author country (Ho, 2014a). In multi-corresponding author articles, all the corresponding authors, institutions, and countries were considered (Al-Moraissi et al., 2023). Articles with corresponding authors in SCI-EXPANDED, that had only address but not affiliation names were checked out and the addresses were changed to be affiliation names (Al-Moraissi et al., 2023).

Affiliations in England, Scotland, North Ireland (Northern Ireland), and Wales were reclassified as being from the United Kingdom (UK) (Chiu and Ho, 2005). Affiliations in Hong Kong before to 1997 were reclassified as in China (Fu and Ho, 2013).

Publications were assessed using following citation indicators:

$C_{year}$: the number of citations from Web of Science Core Collection in a year (e.g. $C_{2022}$ describes citation count in 2022) (Ho, 2012).
$TC_{year}$: the total number of citations from Web of Science Core Collection received since publication year till the end of the most recent year (2022 in this study, $TC_{2022}$) (Wang et al., 2011).
$CPP_{year}$: average number of citations per publication ($CPP_{2022} = TC_{2022}/TP$), $TP$: total number of publications (Ho, 2013).
Six publication indicators were applied to evaluate publication performance of countries and institutions (Hsu and Ho, 2014):
$TP$: total number of articles
$IP$: number of single-country articles ($IP_C$) or number of single-institution articles ($IP_I$)
$CP$: number of internationally collaborative articles ($CP_C$) or number of inter-institutionally collaborative articles ($CP_I$)
$FP$: number of first-author articles
$RP$: number of corresponding-author articles
$SP$: number of single-author articles

Six citation indicators ($CPP_{2022}$) related to the six publication indicators were also applied to evaluate the publication impact on countries and institutions (Ho and Mukul, 2021).
Ho proposed an indicator, the $Y$-index is related to the number of first-author publications ($FP$) and corresponding-author publications ($RP$). The $Y$-index combines two parameters ($j$, $h$) to evaluate the publication potential and contribution characteristics into one index. In the last decade, the index has been applied for comparative authors in a wide range of research topics, for example topics in Social Science Citation Index (SSCI) (Ho, 2014a; Hsu et al., 2020) and Science Citation Index Expanded (SSCI-EXPANDED) (Ho, 2014b; Ho and Shekofteh, 2021). The *Y*-index is defined as (Ho, 2012;2014b):

*Y*-index ($j$, $h$)

where $j$ is a constant related to the publication potential, the sum of the first-author articles and the corresponding-author articles; and $h$ is a constant related to the publication characteristics, polar angle about the proportion of *RP* to *FP*. The greater the value of $j$, the more the first- and corresponding-author contributes to the articles.

$h = \pi/2$, indicates an author that has only published corresponding-author articles, $j$ is the number of corresponding-author articles;

$\pi/2 > h > \pi/4$ indicates that an author has more corresponding-author articles than first-author articles (*FP* > 0);
$h = \pi/4$ indicates that an author has the same number of first- and corresponding-author articles (*FP* > 0 and *RP* > 0);
$\pi/4 > h > 0$ indicates an author with more first-author articles than corresponding-author articles (*RP* > 0);
$h = 0$, indicates that an author has only published first-author articles, $j$ is the number of first-author articles.

## RESULTS AND DISCUSSION

### Characteristics of document types

The characteristics of document type based on their average number of citations per publication ($CPP_{year} = TC_{year}/TP$) and the average number of authors per publication ($APP = AU/TP$) as basic information of document type (Monge-Nájera and Ho, 2017). A total of 3,044 highly cited documents containing search keywords in their 'font page' in SCI-EXPANDED were found among 12 document types which are detailed in Table 1. This publication count includes 2,394 articles (79% of 3,044 documents) with an *APP* of 5.7. The document type of bibliography with one document had the greatest $CPP_{2022}$ value of 350 which was attributed to the only bibliography entitled "Metaheuristics: A bibliography" (Osman and Laporte, 1996) with a $TC_{2022}$ of 350.

The $CPP_{2022}$ of the document type of reviews was found to be 1.2 times of articles. A total of 604 reviews were published in widely 380 journals, mainly in the *Renewable & Sustainable Energy Reviews* (30 reviews; 5.0% of 604 review). There were 58 classic publications with $TC_{2022}$ of 1,000 or more (Long et al., 2014) in highly cited artificial intelligence related study including 41 classic articles, 17 classic reviews, and two classic proceedings papers. It was point out that documents could be categorized in two document types in Web of Science Core Collection, for example, 156 proceedings papers, five book chapters, one data paper, and one publication with expression of concern were also classified in document type of articles. Therefore, cumulative percentages exceed 100% in Table 1 (Usman and Ho, 2020).

Contributions of various document types are different. Generally, articles contain introduction, methods, results, discussion, and conclusion, were chosen for further analyses (Ho and Mukul, 2021). A total of 2,394 articles were presented in two different languages. Article entitled "Neuropsychiatric inventory. The psychometric properties of its adaptation to Spanish" (Vilalta-Franch et al., 1999) was the only published in Spanish with a $TC_{2022}$ of 107.

Table 1. Citations and authors according to the document type.

| Document type | TP | % | AU | APP | $TC_{2022}$ | $CPP_{2022}$ |
|---|---|---|---|---|---|---|
| Article | 2,394 | 79 | 13,591 | 5.7 | 548,886 | 229 |
| Review | 604 | 20 | 2,712 | 4.5 | 165,781 | 274 |
| Proceedings paper | 156 | 5.1 | 612 | 3.9 | 337,95 | 217 |
| Editorial material | 34 | 1.1 | 201 | 5.9 | 62,78 | 185 |
| Book chapter | 8 | 0.26 | 23 | 2.9 | 1,597 | 200 |
| Note | 4 | 0.13 | 21 | 5.3 | 8,56 | 214 |
| Letter | 3 | 0.10 | 36 | 12 | 441 | 147 |
| News item | 3 | 0.10 | 3 | 1.0 | 543 | 181 |
| Bibliography | 1 | 0.033 | 2 | 2.0 | 350 | 350 |
| Data paper | 1 | 0.033 | 11 | 11 | 117 | 117 |
| Publication with expression of concern | 1 | 0.033 | 18 | 18 | 102 | 102 |
| Reprint | 1 | 0.033 | 5 | 5.0 | 109 | 109 |

*TP*: number of publications; *AU*: number of authors; *APP*: average number of authors per publication; $TC_{2022}$: the total number of citations from Web of Science Core Collection since publication year to the end of 2022; $CPP_{2022}$: average number of citations per publication ($TC_{2022}/TP$).

## Characteristics of publication outputs

A study was conducted to examine the relationship between the annual number of highly cited articles (*TP*) and their $CPP_{year}$ by year, aiming to analyze the development trends and impacts of articles on various research topics such as multiple sclerosis (Ho and Shekofteh, 2021) and Phosphoinositide 3-Kinase (Ho and Hartley, 2020), as well as in the Web of Science categories of ophthalmology (Ho et al., 2023), anesthesiology (Juang et al., 2021), emergency medicine (Ho, 2021), and dentistry, oral surgery, and medicine (Yeung and Ho, 2019) within the SCI-EXPANDED database. Figure 1 illustrates the distribution of highly cited articles in the field of artificial intelligence. The year 2017 stands out with 110 articles, achieving the highest $CPP_{2022}$ value of 351. This notable performance can be attributed to two of the top ten most frequently cited articles by Esteva et al. (2017) and Silver et al. (2017), which ranked 3[rd] and 7[th], respectively, with $TC_{2022}$ values of 5,177 and 3,678.

Additionally, in 1994, 1995, 2020, and 2016, there were also higher $CPP_{2022}$ values of 349, 326, 322, and 318, respectively. Among the years analyzed, the year 2019 saw the publication of the highest number of highly cited articles, with a total of 188. Interestingly, the distribution pattern depicted in Fig. 1 differs from that of other research fields, as reported in studies by Ho et al. (2023), Juang et al. (2021), Ho (2021), and Yeung and Ho (2019). Of interest is the relatively rapid ascent of highly cited artificial intelligence articles, as it took only four years to reach its peak, highlighting the high level of activity in this research field. This timeframe stands in contrast to other fields, which typically require around ten years to reach their respective peaks. These findings underscore the significant and dynamic nature of artificial intelligence research.

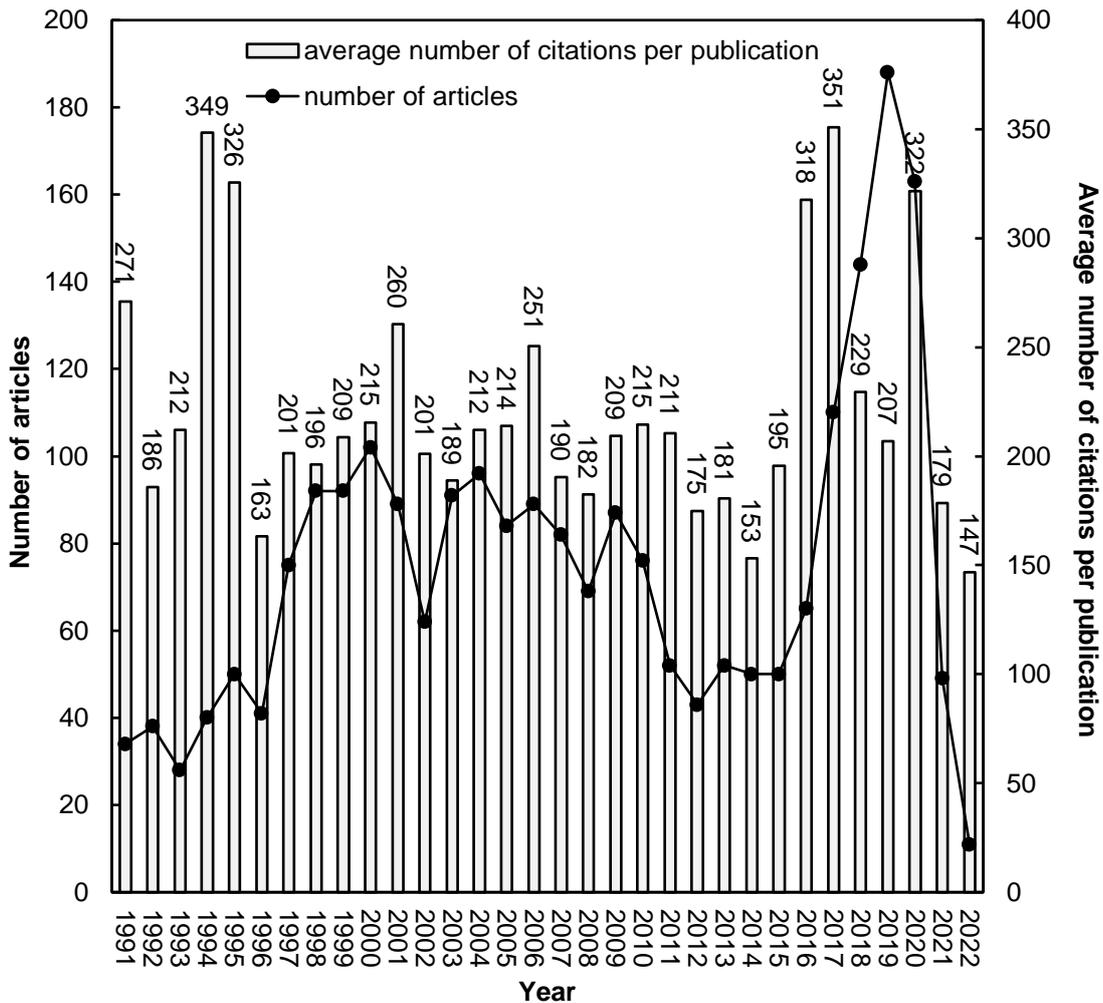

Figure 1. Number of highly cited artificial intelligence articles and their average number of citations per publication by year.

# Web of Science Category and Journal

In 2022, Journal Citation Reports (JCR) indexed 9,510 journals with citation references across 178 Web of Science categories in SCI-EXPANDED. Total 970 journals published highly cited artificial intelligence articles in widely 163 Web of Science categories in SCI-EXPANDED mainly in artificial intelligence computer science with 257 articles (11% of 2,394 articles) and electrical and electronic engineering with 255 articles (11%). In 2021, Ho proposed the characteristics of the journals based on their average number of citations per publication ($CPP_{year}$) and the average number of authors per publication ($APP$) as basic information of the journals in a research topic (Ho, 2021). Table 2 shows the top 10 most productive journals with $IF_{2022}$, $CPP_{2022}$, and $APP$. The *Artificial Intelligence* ($IF_{2022}$ = 14.4) published the most 39 highly cited articles which represented 1.6% of 2,394 articles. Compare to the top 10 productive journals in Table 2, highly cited artificial intelligence articles published in the *Nature* ($IF_{2022}$ = 64.8) had the greatest $CPP_{2022}$ of 910. In contrast, articles in the *Journal of Dairy Science* ($IF_{2022}$ = 3.5) had only 157. The $APP$ ranged from 14 in the *Journal of Clinical Oncology* to 2.3 in the *Artificial Intelligence*. According to $IF_{2022}$, the top six journals have an $IF_{2022}$ of more than 100 were the *CA-A Cancer Journal for Clinicians* ($IF_{2022}$ = 254.7) with two articles, the *Lancet* ($IF_{2022}$ = 168.9) with four articles, the *New England Journal of Medicine* ($IF_{2022}$ = 158.5) with one article, and the *JAMA-Journal of the American Medical Association* ($IF_{2022}$ = 120.7) with two articles, the *Nature Reviews Drug Discovery* ($IF_{2022}$ = 120.1) with one article, and the *BMJ-British Medical Journal* ($IF_{2022}$ = 105.7) with one article. The *CA-A Cancer Journal for Clinicians* ranked not only the top in 241 journals classified in the category of oncology but also the top in 9,510 journals in the SCI-EXPANDED. The *Lancet*, the *New England Journal of Medicine*, the *JAMA-Journal of the American Medical Association*, and the *BMJ-British Medical Journal* ranked the top four in 168 journals classified in the category of general and internal medicine. The *Nature Reviews Drug Discovery* ranked not only the top in 156 journals in the category of biotechnology and applied microbiology but also the top in 277 journals in the category of pharmacology and pharmacy.

Table 2. The top 10 most productive journals.

| Journal | $TP$ (%) | $IF_{2022}$ | $APP$ | $CPP_{2022}$ | Web of Science category |
|---|---|---|---|---|---|
| Artificial Intelligence | 39 (1.6) | 14.4 | 2.3 | 351 | artificial intelligence computer science |
| Proceedings of the National Academy of Sciences of the United States of America | 35 (1.5) | 11.1 | 6.5 | 420 | multidisciplinary sciences |
| Theriogenology | 35 (1.5) | 2.8 | 4.0 | 172 | reproductive biology veterinary sciences |
| Journal of Dairy Science | 33 (1.4) | 3.5 | 5.2 | 157 | dairy and animal science agriculture food science and technology |
| Nature | 33 (1.4) | 64.8 | 9.3 | 910 | multidisciplinary sciences |
| IEEE Access | 32 (1.3) | 3.9 | 4.7 | 232 | information systems computer science electrical and electronic engineering telecommunications |
| Expert Systems with Applications | 27 (1.1) | 8.5 | 2.9 | 188 | artificial intelligence computer science electrical and electronic engineering operations research and management science |
| Journal of Biological Chemistry | 20 (0.84) | 4.8 | 6.5 | 182 | biochemistry and molecular biology |
| Animal Reproduction Science | 19 (0.79) | 2.2 | 4.4 | 181 | dairy and animal science agriculture reproductive biology veterinary sciences |
| Journal of Clinical Oncology | 19 (0.79) | 45.3 | 14 | 290 | oncology |

*TP*: total number of articles; %: percentage of articles in all highly cited artificial intelligence articles; *IF*$_{2022}$: journal impact factor in 2022; *APP*: average number of authors per article; *CPP*$_{2022}$: average number of citations per paper (*TC*$_{2022}$/*TP*).

## Publication performances: countries and institutions

It is widely recognized that two authors: first and the corresponding authors are considered as the most contributed authors in a research article (Riesenberg and Lundberg, 1990). There were three highly cited artificial intelligence articles (0.13% of 2,394 articles) without affiliations in SCI-EXPANDED. A total of 2,391 highly cited articles were published by authors affiliated from 80 countries including 1,641 single-country articles (69% of 2,391 articles) published by authors from 58 countries with a *CPP*$_{2022}$ of 218 and 750 internationally collaborative articles (31%) published by authors from 78 countries with a *CPP*$_{2022}$ of 253. The results demonstrated that internationally collaborative raised citations in the highly cited artificial intelligence research. Six publication indicators and the six related citation indicators (*CPP*$_{2022}$) (Ho and Mukul, 2021) were applied to compare the top 10 productive countries (Table 3). The USA dominated in the six publication indicators with a *TP* of 1,066 highly cited articles (45% of 2,391 highly cited articles), an *IP*$_C$ of 666 articles (41% of 1,641 highly cited single-country articles), a *CP*$_C$ of 400 articles (53% of 750 highly cited internationally collaborative articles), an *FP* of 836 articles (35% of 2,391 highly cited first-author articles), an *RP* of 856 articles (36% of 2,377 highly cited corresponding-author articles), and an *SP* of 88 articles (51% of 174 highly cited single-author articles). Compare to the top 10 productive countries in Table 3, Canada with a *TP* of 152 articles and an *CP*$_C$ of 92 articles the greatest *CPP*$_{2022}$ of 348 and 461 for the *TP* and the *CP*$_C$ respectively. The UK with an *IP*$_C$ of 138 articles, an *FP* of 206 articles, and an *RP* of 214 articles the greatest *CPP*$_{2022}$ of 284, 300, and 302 for the *IP*$_C$, *FP*, and *RP* respectively. Australia with an *SP* of four articles had the greatest *CPP*$_{2022}$ of 409 for the *SP*.

At the institutional level, the determined institution of the corresponding author might be a home base of the study or origin of the paper (Ho, 2012). Concerning institutions, 895 highly cited artificial intelligence articles (37% of 2,391 articles) originated from single institutions with a *CPP*$_{2022}$ of 213 while 1,496 articles (63%) were institutional collaborations with a *CPP*$_{2022}$ of 238. The results showed that collaborations races citations. The top 15 productive institutions and their characteristics are presented in Table 4. Ten of the top 15 productive institutions were in the USA, two in China, two in the UK, and one in Canada. The Chinese Academy of Sciences in China ranked the top with a *TP* of 63 highly cited articles (2.6% of 2,391 highly cited articles), a $CP_I$ of 61 articles (4.1% of 1,496 inter-institutionally collaborative articles), an *FP* of 31 articles (1.3% of 2,391 first-author articles), and an *RP* of 44 articles (1.9% of 2,347 corresponding-author articles). However, the Chinese Academy of Sciences had a lower $CPP_{2022}$ for all kind of articles. Furthermore, the bias emerged due to the presence of branches of the Chinese Academy of Sciences across various cities (Li et al., 2009). Currently, the institution's publications are consolidated under a single heading, but if they were categorized by branches, it would lead to disparate rankings (Li et al., 2009). Except for the Chinese Academy of Sciences, the Massachusetts Institute of Technology (MIT) dominated in five of the six publication indicators with a *TP* of 44 highly cited articles (1.8% of 2,391 highly cited articles), an $IP_I$ of 15 articles (1.7% of 895 single-institution articles), an *FP* of 29 articles (1.2% of 2,391 first-author articles), an *RP* of 29 articles (1.2% of 2,347 corresponding-author articles), and an *SP* of five articles (2.9% of 174 single-author articles). The Stanford University in USA dominated with an $CP_I$ of 34 articles (2.3% of 1,496 inter-institutionally collaborative articles). Compared to the top 15 productive institutions in Table 4, the Stanford University in USA with *TP* of 42 articles, an $CP_C$ of 34 articles, an *FP* of 21 articles, an *RP* of 22 articles, and an *SP* of two articles had the greatest $CPP_{2022}$ of 493, 423, 763, 690 and 1,602 for *TP*, $CP_C$, *FP*, *RP*, and *SP*. The University College London (UCL) in the UK with an $IP_I$ of one article had the greatest $CPP_{2022}$ of 2,513 for $IP_I$.

Table 3. Top 10 productive countries.

| Country | TP | TP | | $IP_C$ | | $CP_C$ | | FP | | RP | | SP | |
|---|---|---|---|---|---|---|---|---|---|---|---|---|---|
| | | R (%) | $CPP_{2022}$ | R (%) | $CPP_{2022}$ | R (%) | $CPP_{2022}$ | R (%) | $CPP_{2022}$ | R (%) | $CPP_{2022}$ | R (%) | $CPP_{2022}$ |
| USA | 1,066 | 1 (45) | 268 | 1 (41) | 243 | 1 (53) | 309 | 1 (35) | 271 | 1 (36) | 268 | 1 (51) | 343 |
| China | 376 | 2 (16) | 212 | 2 (11) | 203 | 2 (27) | 220 | 2 (12) | 206 | 2 (13) | 206 | 8 (1.7) | 245 |
| UK | 312 | 3 (13) | 282 | 3 (8.4) | 284 | 3 (23) | 280 | 3 (8.6) | 300 | 3 (9.0) | 302 | 2 (10) | 271 |
| Germany | 166 | 4 (6.9) | 200 | 6 (3.2) | 176 | 4 (15) | 211 | 5 (3.9) | 192 | 4 (4.2) | 190 | 6 (2.3) | 192 |
| Canada | 152 | 5 (6.4) | 348 | 5 (3.7) | 175 | 5 (12) | 461 | 4 (3.9) | 223 | 5 (4.0) | 225 | 4 (2.9) | 154 |
| France | 126 | 6 (5.3) | 219 | 7 (3.0) | 174 | 6 (10) | 249 | 7 (3.2) | 175 | 7 (3.3) | 181 | 3 (5.2) | 151 |
| Japan | 122 | 7 (5.1) | 182 | 4 (4.0) | 165 | 9 (7.6) | 202 | 6 (3.8) | 171 | 6 (3.7) | 174 | 8 (1.7) | 179 |
| Australia | 115 | 8 (4.8) | 207 | 8 (2.7) | 200 | 7 (9.3) | 211 | 8 (2.7) | 211 | 8 (2.9) | 218 | 6 (2.3) | 409 |
| Italy | 102 | 9 (4.3) | 198 | 9 (2.6) | 166 | 8 (8.0) | 220 | 9 (2.3) | 185 | 9 (2.4) | 182 | 4 (2.9) | 205 |
| Netherlands | 73 | 10 (3.1) | 254 | 10 (1.2) | 154 | 10 (7.2) | 290 | 14 (1.4) | 161 | 14 (1.4) | 160 | 13 (1.1) | 132 |

*TP*: number of total articles; *TP R* (%): total number of articles and the percentage of total articles; $IP_C$ *R* (%): rank and percentage of single-country articles in all single-country articles; $CP_C$ *R* (%): rank and percentage of internationally collaborative articles in all internationally collaborative articles; *FP R* (%): rank and the percentage of first-author articles in all first-author articles; *RP R* (%): rank and the percentage of corresponding-author articles in all corresponding-author articles; *SP R* (%): rank and the percentage of first-author articles in all first-author articles; $CPP_{2022}$: average number of citations per publication ($CPP_{2022} = TC_{2022}/TP$); N/A: not available.

Table 4. Top 15 productive institutions.

| Institution | TP | TP | | IPc | | CPc | | FP | | RP | | SP | |
|---|---|---|---|---|---|---|---|---|---|---|---|---|---|
| | | R (%) | CPP2022 | R (%) | CPP2022 | R (%) | CPP2022 | R (%) | CPP2022 | R (%) | CPP2022 | R (%) | CPP2022 |
| Chinese Acad Sci, China | 63 | 1 (2.6) | 185 | 66 (0.22) | 165 | 1 (4.1) | 186 | 1 (1.3) | 201 | 1 (1.9) | 189 | 10 (1.1) | 287 |
| MIT, USA | 44 | 2 (1.8) | 316 | 1 (1.7) | 319 | 6 (1.9) | 315 | 2 (1.2) | 344 | 2 (1.2) | 324 | 1 (2.9) | 570 |
| Stanford Univ, USA | 42 | 3 (1.8) | 493 | 5 (0.89) | 790 | 2 (2.3) | 423 | 4 (0.88) | 763 | 3 (0.94) | 690 | 10 (1.1) | 1602 |
| Harvard Univ, USA | 40 | 4 (1.7) | 241 | 11 (0.78) | 175 | 3 (2.2) | 254 | 6 (0.84) | 249 | 6 (0.81) | 247 | 3 (1.7) | 210 |
| Univ Oxford, UK | 31 | 5 (1.3) | 334 | 152 (0.11) | 324 | 4 (2.0) | 335 | 20 (0.42) | 215 | 19 (0.51) | 213 | 25 (0.57) | 175 |
| Univ Florida, USA | 31 | 5 (1.3) | 160 | 2 (1.6) | 158 | 21 (1.1) | 162 | 4 (0.88) | 146 | 5 (0.85) | 148 | 10 (1.1) | 148 |
| Tsinghua Univ, China | 31 | 5 (1.3) | 204 | 152 (0.11) | 111 | 4 (2.0) | 207 | 15 (0.50) | 184 | 9 (0.68) | 190 | N/A | N/A |
| Univ Toronto, Canada | 29 | 8 (1.2) | 283 | 18 (0.56) | 190 | 9 (1.6) | 302 | 23 (0.38) | 574 | 24 (0.43) | 533 | 25 (0.57) | 132 |
| Univ Michigan, USA | 29 | 8 (1.2) | 213 | 66 (0.22) | 141 | 7 (1.8) | 218 | 13 (0.59) | 174 | 12 (0.64) | 171 | 25 (0.57) | 267 |
| Univ Calif San Francisco, USA | 28 | 10 (1.2) | 279 | 5 (0.89) | 237 | 15 (1.3) | 296 | 12 (0.63) | 229 | 12 (0.64) | 233 | 10 (1.1) | 647 |
| Univ Calif Berkeley, USA | 28 | 10 (1.2) | 202 | 18 (0.56) | 192 | 11 (1.5) | 204 | 11 (0.67) | 219 | 15 (0.6) | 219 | 3 (1.7) | 214 |
| UCL, UK | 28 | 10 (1.2) | 354 | 152 (0.11) | 2513 | 7 (1.8) | 274 | 68 (0.21) | 653 | 46 (0.30) | 679 | N/A | N/A |
| Princeton Univ, USA | 28 | 10 (1.2) | 317 | 3 (1.5) | 235 | 30 (1.0) | 387 | 3 (0.92) | 305 | 3 (0.94) | 312 | N/A | N/A |
| Univ Maryland, USA | 26 | 14 (1.1) | 215 | 14 (0.67) | 169 | 15 (1.3) | 229 | 13 (0.59) | 154 | 15 (0.60) | 159 | 10 (1.1) | 196 |
| Texas A&M Univ, USA | 26 | 14 (1.1) | 167 | 18 (0.56) | 157 | 12 (1.4) | 170 | 20 (0.42) | 163 | 22 (0.47) | 187 | N/A | N/A |

$TP$: total number of articles; $TP$ $R$ (%): total number of articles and percentage of total articles; $IP_1$ $R$ (%): rank and percentage of single-institution articles in all single-institution articles; $CP_1$ $R$ (%): rank and percentage of inter-institutionally collaborative articles in all inter-institutionally collaborative articles; $FP$ $R$ (%): rank and percentage of first-author articles in all first-author articles; $RP$ $R$ (%): rank and percentage of corresponding-author articles in all corresponding-author articles; $SP$ $R$ (%): rank and percentage of single-author articles in all single-author articles; $CPP_{2022}$: average number of citations per publication ($CPP_{2022} = TC_{2022}/TP$); N/A: not available.

## Publication performances: authors

For articles related to highly cited artificial intelligence, the $APP$ was 5.7 whereas the maximum number of authors was 225 in one article (Sarwar et al., 2010). Of the 2,391 articles with author information, 57% articles were published by groups of two to five authors, including 400 highly cited articles (17% of 2,391 highly cited articles), 357 (15%), 342 (14%), and 272 (11%) were written by groups of 3, 2, 4, and 5 authors, respectively. Table 5 listed the top 12 productive authors with 10 highly cited artificial intelligence articles or more. B.L. Bassler was the most productive author with 18 highly cited articles including the most 14 corresponding-author articles. J.E.P. Santos with 15 articles, published the most five first-author articles. In addition, M. Chen with nine articles, also published five first-author articles. Compare to the 12 productive authors, D. Hassabis had not only the greatest $CPP_{2022}$ of 1,538 for all highly cited articles but also the greatest $CPP_{2022}$ of 4,091 for corresponding-author articles. Z.L. Wang had the greatest $CPP_{2022}$ of 287 for first-author articles. Six of the 12 productive authors including J.E.P. Santos, B.L. Bassler, Z.L. Wang, T.K. Wood, Y. Zhang, and M.C. Wiltbank were found to be the top 12 publication potential authors as evaluated by $Y$-index. In the total of 2,292 highly cited artificial intelligence articles (96% of 2,394 highly cited articles) had both first and corresponding authors information in SCI-EXPANDED, were extensively investigated based on the $Y$-index. The 2,292 highly cited artificial intelligence related articles were contributed by 11,079 authors in which 8,136 authors (73% of 11,079 authors) had no first- and no corresponding-author articles with $Y$-index (0, 0); 837 (7.6%) authors published only corresponding-author articles with an $h$ of $\pi/2$; 59 (0.53%) authors published more corresponding-author articles than first-author articles with $\pi/2 > h > \pi/4$ ($FP > 0$); 1,226 (11%) authors published the same number of first- and corresponding-author articles with an $h$ of $\pi/4$ ($FP > 0$ and $RP > 0$); 24 (0.22%) authors published more first-author articles than corresponding-author articles with $\pi/4 > h > 0$ ($RP > 0$); and 797 (7.2%) authors published only first-author articles with an $h$ of 0.

In the polar coordinates (Fig. 2), the distribution of the $Y$-index ($j$, $h$) of the leading 36 potential authors in highly cited artificial intelligence articles with $j \geq 5$ was demonstrated. Every point has a coordinate $Y$-index ($j$, $h$) that could symbolize a single author or multiple authors, for example, H.J.W.L. Aerts, J.W. Gu, and W.E. Bentley with $Y$-index (5, $\pi/2$) and F. Tao, G.H. Recanzone, P. Leitao, F. Lopez-Gatius, L. Floridi, J.H. Lee, and P. Liu with $Y$-index (6, $\pi/4$). J.E.P. Santos with $Y$-index (17, 1.176) had the greatest publication potential in highly cited articles, followed by B.L. Bassler (14, $\pi$ /2). T.K. Wood (8, $\pi/2$), Y. Zhang (8, 1.429), T.J. Brinker and D. Silver (8, 1.030), D.E. Swayne, D.T. Bui, and W. Chen (8, $\pi/4$), and M. Chen (8, 0.5404) all had the same $j$ of 8. These authors are located on the same curve ($j = 8$) in Fig. 2, indicating that they had the same publication potential in highly cited articles with a $j$ of 8 but different publication characteristics (Ho and Hartley, 2016b). Wood only published corresponding-author articles with an $h$ of $\pi/2$. Zhang had the highest ratio of corresponding-author articles to first-author articles with an $h$ of 1.429; followed by Brinker and Silver with an $h$ of 1.030; and Swayne, Bui, and Chen with an $h$ of $\pi/4$. However, Chen published more first-author articles than corresponding-author articles with an $h$ of 0.5404. Similar situation for authors located on $j$ of 7, 6, and 5 was also found.

D.E. Swayne (8, $\pi/4$), D.T. Bui (8, $\pi/4$), W. Chen (8, $\pi/4$), F. Tao (6, $\pi/4$), G.H. Recanzone (6, $\pi/4$), P. Leitao (6, $\pi/4$), F. Lopez-Gatius (6, $\pi/4$), L. Floridi (6, $\pi/4$), J.H. Lee (6, $\pi/4$), P. Liu (6, $\pi/4$) are located on the diagonal line ($h = \pi/4$) indicating that they had the same publication characteristics but different publication potential. Swayne, Bui, and Chen had the greatest publication potential with a $j$ of 8 followed by Tao, Recanzone, Leitao, Lopez-Gatius Floridi, Lee, and Liu with a $j$ of 6. Similarly, B.L. Bassler (14, $\pi/2$), T.K. Wood (8, $\pi/2$), H.J.W.L. Aerts (5, $\pi/2$), J.W. Gu (5, $\pi/2$), and W.E. Bentley (5, $\pi/2$) are located on the straight line (y-axis) ($h = \pi/2$). Bassler had the highest publication potential with a $j$ of 14 followed by Wood with a $j$ of 8, Aerts, Gu, and Bentley with a $j$ of 5. The location on the graph along with one of the curves or along a line from the origin represents different families of author publication potential or publication characteristics, respectively. A potential for bias in the analysis of authorship might attributes to different authors having the same name, or the same author using different names over time (Chiu and Ho, 2007).

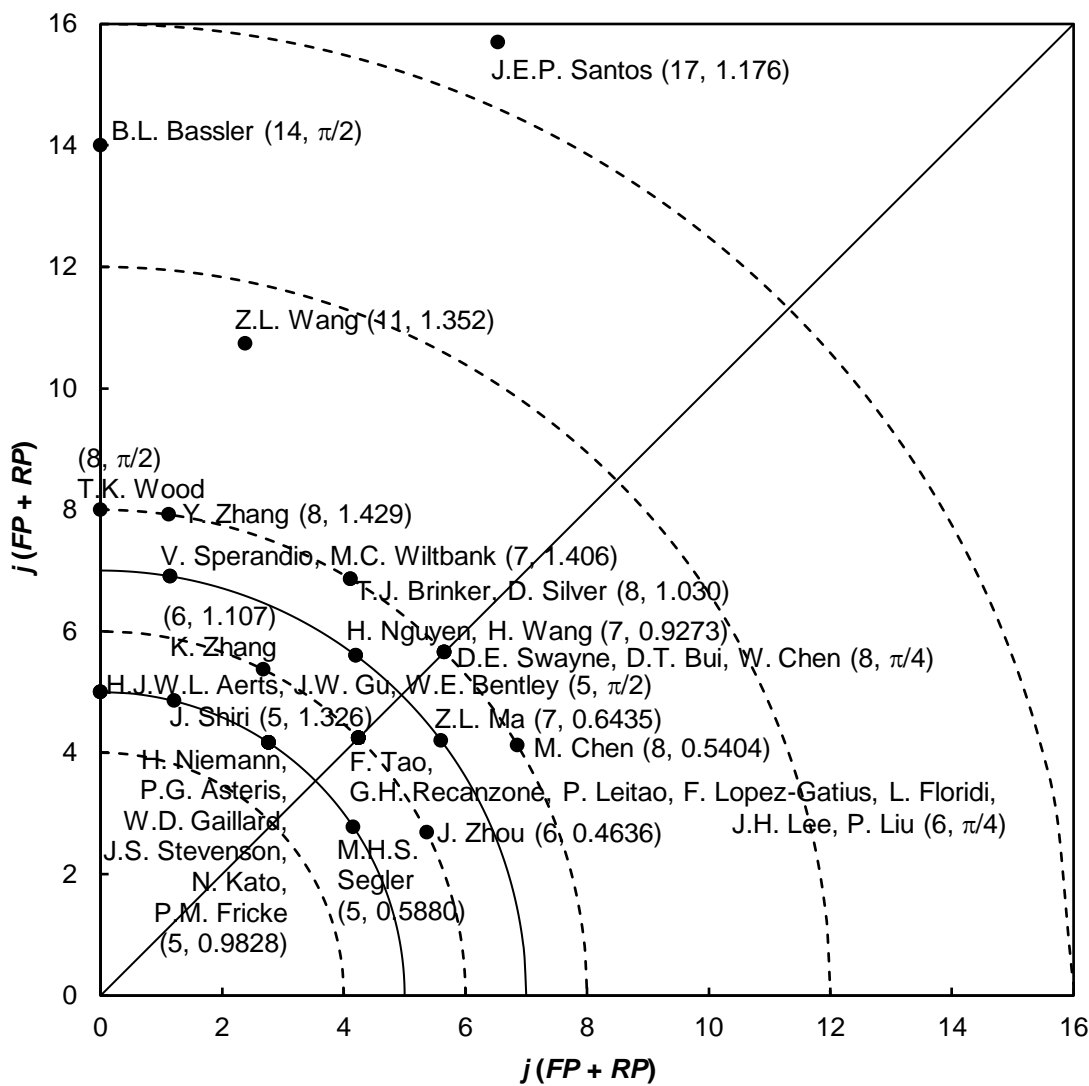

Figure 2. Top 36 authors with *Y*-index ($j \geq 5$)

## The nine classic articles in highly cited artificial intelligence research

Publications with $TC_{year}$ of 1,000 or more was named classic publications (Long et al., 2014; Tantengco and Ho, 2021). There were 41 classic articles in artificial intelligence research. Total citations are updated from time to time on the Web of Science Core Collection. To improve bibliometric study, the total number of citations from the Web of Science Core Collection since publication year to the end of the most recent year of 2022 ($TC_{2022}$) was applied to improve the bias using data from the database directly (Wang et al., 2011). A total of 275 articles (0.11% of 2,394 articles), 2,105 articles (88% of 2,381 articles with abstract in SCI-EXPANDED), and 435 articles (27% of 1,582 articles with author keywords in SCI-EXPANDED) contain search keywords in their title, abstract, and author keywords respectively. An article entitled "Generative Adversarial Networks" (Goodfellow et al., 2020) was the most frequently cited with a $TC_{2022}$ of 20,068. The authors mentioned that "Generative adversarial networks are a kind of artificial intelligence algorithm designed to solve the generative modeling problem" in the abstract. Four, 31, and 7 of the 41 classic articles contain search keywords in their title, abstract, and author keywords respectively. Thirty-one of the articles mentioned the search keywords solely in their abstract, less focusing on artificial intelligence. The remaining nine articles containing search keywords in their title or author keywords were listed in Table 6. Citations of a highly cited article is not always high (Ho, 2014b). It is necessary to understand citation history of a classic article. The citation histories of the nine classic artificial intelligence articles are shown in Fig. 3.

Table 6 and Fig. 3 are related. Table 6 provides the nine classic articles discussed above. They are ordered according to the number of citations since publication ($TC_{2022}$) but the position of the number of citations in the year 2022 ($C_{2022}$) is also given. In the same vein, Fig. 3 shows the citation life cycle of the nine classic articles.
According to the articles displayed in Table 6, the number of citations received in the most recent year of 2022 ($C_{2022}$) represents a high percentage of the total citations, indicating the interest and impact of the analysed articles. This can be seen visually in Fig. 3.
Article by Kirkpatricka et al. (2017) not only ranked top with $TC_{2022}$ of 1,583 but also ranked in good places in $C_{2022}$.

Country representation is dominated by the UK (with 3 publications) followed by the USA (2), the remaining articles are signed by a single country, coming from Spain, France, Belgium, Morocco and Australia. Seven articles (in the first seven positions of the ranking) are signed by more than one author, while the other two articles (positions eight and nine) are written by a single author.
The subject matter or scientific areas of the nine articles are disparate, although there are two main groups that stand out. On the one hand, two articles focus on artificial neural networks (publications by Kirkpatricka et al. (2017) and Gardner and Dorling (1998). Another recurring theme is that of authors Browne et al. (2012) and Lundberg et al. (2020) where they write about tree-based machine learning models. The rest of the articles have a unique theme, where AI aims to improve medicine, social sciences, image or video detectors, etc.

The nine articles are represented in Fig. 3 where the number of citations is related to the year of publication. Here two temporal spaces stand out, one with a single publication in 1998 (by the authors Gardner and Dorling with very few citations until 2015) and another period from 2011 when the rest of the articles begin to be published. In this space, the vast majority of publications offer an almost exponential growth in citations annually.
In relation to the number of citations, it can be seen in Fig. 3 that the most current articles offer a very pronounced upward trend, for example, the article by the Spanish Arrieta et al. or the American

Lundberg et al. from 2020. There is only one article with a downward trend in terms of the number of citations per year, the one written by the Belgian authors Barnich and Van Droogenbroeck in 2011.

Table 5. Top 12 productive authors with 10 highly cited articles or more

| Author | *TP* | | *FP* | | *RP* | | *h* | rank (*j*) |
|--------|------|--|------|--|------|--|----|------------|
| | rank (*TP*) | $CPP_{2022}$ | rank (*FP*) | $CPP_{2022}$ | rank (*RP*) | $CPP_{2022}$ | | |
| B.L. Bassler | 1 (18) | 350 | N/A | N/A | 1 (14) | 314 | $\pi/2$ | 2 (14) |
| J.E.P. Santos | 2 (15) | 179 | 1 (5) | 225 | 2 (12) | 190 | 1.176 | 1 (17) |
| M.C. Wiltbank | 3 (12) | 267 | 152 (1) | 136 | 6 (6) | 196 | 1.406 | 12 (7) |
| Z.L. Wang | 3 (12) | 260 | 35 (2) | 287 | 3 (9) | 193 | 1.352 | 3 (11) |
| J. Zhang | 5 (11) | 182 | 152 (1) | 148 | 211 (1) | 148 | $\pi/4$ | 184 (2) |
| Y. Liu | 5 (11) | 458 | 3 (4) | 137 | N/A | N/A | 0 | 37 (4) |
| J. Li | 5 (11) | 142 | 152 (1) | 166 | 59 (2) | 114 | 1.107 | 109 (3) |
| D. Hassabis | 5 (11) | 1538 | N/A | N/A | 59 (2) | 4091 | $\pi/2$ | 184 (2) |
| Y. Zhang | 9 (10) | 171 | 152 (1) | 153 | 5 (7) | 136 | 1.429 | 4 (8) |
| J. Zhou | 9 (10) | 134 | 3 (4) | 147 | 59 (2) | 105 | 0.4636 | 17 (6) |
| J. Wang | 9 (10) | 191 | N/A | N/A | 59 (2) | 232 | $\pi/2$ | 184 (2) |
| T.K. Wood | 9 (10) | 200 | N/A | N/A | 4 (8) | 211 | $\pi/2$ | 4 (8) |

*TP*: total number of articles; *FP*: number of first-author articles; *RP*: number of corresponding-author articles; $CPP_{2022}$: average number of citations per publication ($CPP_{2022} = TC_{2022}/TP$); *j*: *Y*-index constant related to the publication potential; *h*: *Y*-index constant related to the publication characteristics; N/A: not available.

Table 6. Nine classic articles with $TC_{2022}$ of 1,000 or more in artificial intelligence research

| Rank ($TC_{2022}$) | Rank ($C_{2022}$) | Title | Country | Reference |
|--------------------|-------------------|-------|---------|-----------|
| 18 (1,583) | 10 (559) | Overcoming catastrophic forgetting in neural networks | UK | Kirkpatricka et al. (2017) [78] |
| 19 (1,551) | 33 (256) | Artificial neural networks (the multilayer perceptron): A review of applications in the atmospheric sciences | UK | Gardner and Dorling (1998) [79] |
| 22 (1,504) | 6 (813) | Explainable Artificial Intelligence (XAI): Concepts, taxonomies, opportunities and challenges toward responsible AI | Spain, France | Arrieta et al. (2020) [80] |
| 26 (1,292) | 149 (102) | ViBe: A universal background subtraction algorithm for video sequences | Belgium | Barnich and Van Droogenbroeck (2011) [81] |
| 27 (1,279) | 9 (582) | Peeking inside the black-box: A survey on explainable artificial intelligence (XAI) | Morocco | Adadi and Berrada (2018) [82] |
| 28 (1,275) | 56 (187) | A survey of Monte Carlo tree search methods | UK | Browne et al. (2012) [83] |
| 32 (1,164) | 8 (679) | From local explanations to global understanding with explainable AI for trees | USA | Lundberg et al. (2020) [84] |
| 36 (1,112) | 22 (326) | Machine learning in medicine | USA | Deo (2015) [85] |
| 41 (1,044) | 14 (408) | Explanation in artificial intelligence: Insights from the social sciences | Australia | Miller (2019) [86] |

$TC_{2022}$: the total number of citations from Web of Science Core Collection since publication year to the end of 2022; $C_{2022}$: number of citations of an article in 2022 only.

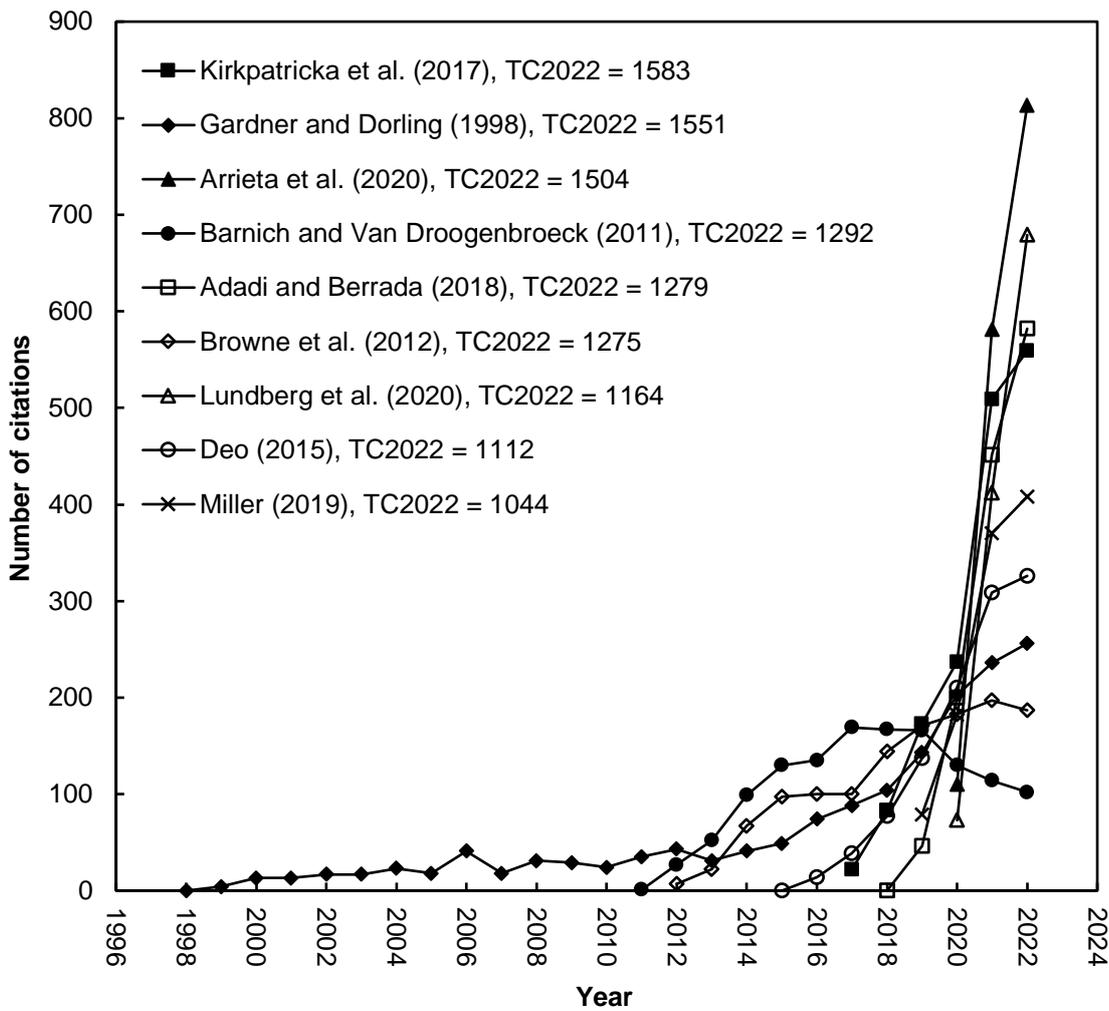

Figure 3. The citation histories of the nine classic articles on artificial intelligence research

CONCLUSION

Between 1991 and 2022, a total of 3,044 highly cited documents on artificial intelligence were published across 12 document types in the SCI-EXPANDED database. The pinnacle of highly cited articles emerged in the year 2019. These influential articles spanned a diverse spectrum of categories, showcasing the breadth of AI research. The United States stood out as the most prolific contributor across all six publication types. The Chinese Academy of Sciences claimed the lead in terms of sheer volume for highly cited articles, albeit with a lower impact. In contrast, Stanford University in the USA had the publication of the most impactful articles with a substantial average number of citations per publication. Notably, B.L. Bassler emerged as the primary contributor of highly cited articles, while J.E.P. Santos exhibited the greatest potential for prolific highly cited articles. A standout piece in this landscape was the 2020 article by Goodfellow et al., which not only had the position of being the most frequently cited but also had the most citations in 2022 as the most influential contribution to artificial intelligence research.